\documentclass[12pt,a4paper]{article}

\usepackage[utf8]{inputenc}
\usepackage[T1]{fontenc}
\usepackage{lmodern}
\usepackage[english]{babel}
\usepackage{csquotes}
\usepackage{microtype}
\usepackage{chngcntr}

\usepackage[margin=2cm]{geometry}
\usepackage{setspace}
\usepackage{amsmath}
\usepackage{amssymb}
\usepackage{amsthm}
\usepackage{mathrsfs}
\usepackage{centernot}
\theoremstyle{definition}
\newtheorem{assumption}{Assumption}
\newtheorem{assumptionA}{Assumption}

\newtheorem{theorem}{Theorem}
\newtheorem{proposition}{Proposition}

\newtheorem{remark}{Remark}
\newcommand{\indep}{\perp\!\!\!\perp}
\newcommand{\noindep}{\centernot{\perp\!\!\!\perp}}

\newcommand{\OR}{\operatorname{OR}}

\usepackage{booktabs}
\usepackage{graphicx}
\usepackage{float}
\usepackage{caption}
\usepackage{subcaption}
\usepackage{multirow}
\usepackage{tikz}
\usetikzlibrary{arrows.meta, positioning, shapes.geometric, calc}
\usepackage[dvipsnames]{xcolor}
\usepackage[backend=bibtex,style=authoryear,maxnames=999,maxcitenames=3]{biblatex}
\renewbibmacro{in:}{}

\usepackage[colorlinks=true,
linkcolor=blue,
citecolor=blue,
urlcolor=blue]{hyperref}

\usepackage{enumitem}  
\usepackage{siunitx}   
\usepackage{xcolor}    
\usepackage{todonotes} 
\usepackage{pdflscape} 
\usepackage{booktabs}
\usepackage{threeparttable}
\usepackage{multirow}
\usepackage{caption}
\usepackage{geometry}

\title{\textbf{Semiparametric Difference-in-Differences Estimation With Missing Not at Random Data: A Shadow Variable Approach}}
\author{
	\small{Junjie Li}\\
	\small{Hitotsubashi University}\\
	\&\\
  \small{Dongyuan Mu}\\
	\small{Hitotsubashi University}\\
}
\date{\today}

\begin{document}
\maketitle

 \begin{abstract}
        This paper considers a semiparametric difference-in-differences (DID) framework for identifying and estimating treatment effects on the treated (ATT) when outcomes are missing not at random (MNAR), and a fully observed shadow variable is available. The shadow variable is assumed to be associated with the outcome evolution but independent of the missingness process, conditional on covariates and the possibly unobserved outcome evolution. We establish the identification conditions, derive the corresponding identification results and estimation algorithm, and evaluate the finite-sample performance of the proposed estimator through simulation studies and a real data application.
    \end{abstract}
    \vspace{1em}\hrule
	\begin{flushleft}
	\end{flushleft}
    \clearpage
  
\section{Introduction}
The difference-in-differences (DID) design is a quasi-experimental method widely used in the social sciences to evaluate causal effects with repeated observations. In its traditional linear form, DID identifies the average treatment effect on the treated (ATT) under the (unconditional) parallel trends assumption (PTA): in the absence of treatment, the average outcomes of the treatment and comparison groups would have followed parallel paths over time. However, this assumption may fail when observed characteristics associated with outcome variation are imbalanced between the treated and comparison groups. To address this issue, \textcite{abadie2005semiparametric} proposed a semiparametric DID estimator that assumes the PTA holds only after conditioning on covariates. Recently, researchers have developed a variety of semiparametric DID estimators. Among them, the doubly robust DID estimator is particularly appealing because it is consistent if either of the working models is correctly specified, but not necessarily both; see, e.g., \textcite{sant2020doubly}, \textcite{chang2020double}, and \textcite{li2025difference}. One main challenge in practice is that survey data are often subject to nonresponse, a problem that tends to be more severe in panel data. A commonly used approach for handling missing data, known as complete-case analysis, discards observations that contain missing values over time. An alternative approach is imputation, which involves imputing or estimating values for the missing outcomes; see Chapter 6 of \textcite{tsiatis2006semiparametric} for further details. Generally, these approaches rely on the missing at random (MAR) assumption, meaning that missingness is ignorable conditional on covariates and treatment.

However, in some practical situations, nonresponse depends on the unobserved outcome itself, i.e., the outcome is missing not at random (MNAR). For example, in evaluating the causal effect of mobile payment usage on household consumption, nonresponse in consumption data is likely to be related to the level of consumption, thereby violating the MAR assumption. To solve this problem, \textcite{shin2024difference} propose a ``bespoke'' instrumental variable (``bespoke'' IV) DID estimand that builds on the idea of \textcite{dukes2022alternative}. The ``bespoke'' IV strategy requires an additional auxiliary variable satisfying restrictions such as relevance to missingness and independence from outcome evolution across groups, thereby enabling identification under nonignorable missingness. However, in small-scale observational datasets, it is often difficult to find an additional variable that satisfies these validity conditions. Hence, exploring an MNAR DID estimator without relying on an auxiliary variable is nontrivial, but it becomes feasible when a \textit{shadow variable} is available. Loosely speaking, a shadow variable is a subset of the covariates (or possibly the entire set) that is associated with the outcome, independent of the missingness process conditional on the outcome and other covariates, and observed for all individuals. The shadow variable was initially called a nonresponse instrument \parencites{d2010new}{wang2014instrumental}{zhao2015semiparametric}; however, to avoid confusion with the literature on instrumental variables for MNAR data \parencites{tchetgen2017general}{sun2018semiparametric}{sun2022semiparametric}, we follow \parencites{miao2016varieties}{zhao2022versatile}{miao2024identification} and term it a \textit{shadow variable}.

The shadow variable approach is useful and feasible in many empirical studies because some covariates may be unrelated to missingness directly but related to the outcome. For example, in a children's mental health study, the teacher's assessment of a student's psychopathology status was treated as the outcome variable and was suspected to be related to missingness because a teacher may be more likely to complete the assessment when the child appears abnormal. In this case, the parent's report on the child can serve as a valid shadow variable (\textcite{zhao2022versatile}). The shadow variable approach can also be used in house price studies. Nonresponse for home prices is likely to be related to the home price itself because homeowners may avoid disclosing a high price that reveals their assets. In this case, construction price can be regarded as a valid shadow variable, since it is related to the current price (\textcite{miao2024identification}).

Although many studies have examined identification and estimation under MNAR using the shadow variable approach, to the best of our knowledge, no study has investigated DID estimation with a shadow variable. Many empirical studies rely on panel data to estimate causal policy effects, and outcomes in the post-treatment period are more likely to be missing than outcomes in the pre-treatment period, with missingness often related to the outcome itself. For example, \textcite{deng2021does} carry out two-period DID analysis on household debt and fertility. The MNAR problem is significant because households with substantial debt increases may be reluctant to disclose their financial situation, making the second-period debt outcome more likely to be missing for those experiencing larger borrowing growth.

In this paper, we use the conditional PTA and extend the ``bespoke'' IV DID estimand to a covariate-adjusted ``bespoke'' IV DID estimand to address nonignorable missingness. We also provide an alternative approach that does not require finding an additional valid variable to construct a DID estimand under MNAR.

\section{Setup}
Our analysis is based on a two-period, two-group structure (treatment and comparison groups). Let $Y_{it}$ represent the realized outcome for unit $i$ at time $t$. We assume that researchers have access to outcome data from both a pre-treatment period ($t = 0$) and a post-treatment period ($t = 1$). 
Let $R_{i}$ denote the missingness indicator for the outcome in the post-treatment period, and assume that there is no missingness in the pre-treatment period. Let $D_{i}$ be a binary treatment indicator and $Y_{it}(d)$, for $d\in\{0,1\}$, denote the potential outcome as in \textcite{rubin1974estimating}. Let $X_i$ denote the pre-treatment covariates. The observable data $\{R_iY_{i1},Y_{i0},R_i,D_i,X_i\}_{i=1}^{n}$ are assumed to be independent and identically distributed. In semiparametric DID estimation, the parameter of interest is the ATT:
\begin{equation*}
\tau = \mathbb{E} [Y_{1}(1)-Y_{1}(0) \mid D=1],
\end{equation*}
where we omit the individual index for ease of notation. In the analysis of fully observed data, researchers typically rely on the following basic assumptions to identify the ATT under the DID design \parencites{heckman1997matching}{abadie2005semiparametric}{sant2020doubly}.

\begin{assumption} [Basic Assumption]\label{basic_assumption}
\end{assumption}
\vspace{-8ex}
\begin{align*}
&\mathrm{(i)} \quad  Y_{t} = D Y_{t}(1) + (1 - D) Y_{t}(0), \  t = 1 \quad & (\text{Consistency}) \\
&\mathrm{(ii)} \quad  Y_{t} = Y_{t}(0), \  t = 0 \quad & (\text{No anticipation of treatment}) \\
&\mathrm{(iii)} \quad  \mathbb{E}[Y_{1}(0) - Y_{0}(0) \mid X, D = 1]
= \mathbb{E}[Y_{1}(0) - Y_{0}(0) \mid X, D = 0] \quad & (\text{Conditional parallel trends})\\
&\mathrm{(iv)} \quad   \text{For some} \  \varepsilon>0, \ \Pr(D=1)> \varepsilon \  \text{and} \  \Pr(D=1|X)\leq 1-\varepsilon \ \text{a.s.} \quad & (\text{Overlap condition})
\end{align*}

Assumption~\ref{basic_assumption} (i) indicates that the realized post-treatment outcome is one of the potential outcomes, and (ii) implies that no unit is exposed to treatment in the pre-treatment period. Assumptions (i) and (ii) can be regarded as variants of the stable unit treatment value assumption (SUTVA). Assumption (iii) is the conditional parallel trends assumption (PTA), which posits that the average conditional outcomes for the treatment and comparison groups would have followed parallel paths in the absence of treatment. This is the key assumption for ATT identification under the DID framework. Assumption (iv) is commonly used in inverse probability weighting estimation and ensures that the denominator probability is well-defined. Based on Assumption~\ref{basic_assumption}, let $\Delta Y= Y_{1} - Y_{0}$. The ATT can be rewritten as
\begin{equation}
  \tau=\mathbb{E}\left[\mathbb{E}[\Delta Y \mid D=1,X] \mid D=1\right]-\mathbb{E}\left[\mathbb{E}[\Delta Y \mid D=0,X] \mid D=1\right] \label{full_identification}
\end{equation}
It is clear that $\mathbb{E}[\Delta Y \mid D=d,X]$, for $d\in\{0,1\}$, is not identifiable from the observed data when outcomes are missing. Next, we introduce the missingness mechanisms under the DID design. The missingness mechanism for outcomes under the DID framework can be expressed as:
\begin{assumption}[Missingness Mechanism] \label{missingness_mechanism} 
\end{assumption} 
\vspace{-8ex} 
\begin{align*} 
  & \mathrm{(i)} \quad R \indep \Delta Y \mid (D, X) \quad &(\text{MAR}) \\
   & \mathrm{(ii)} \quad R \noindep \Delta Y \mid (D, X) \quad &(\text{MNAR}) 
  \end{align*}
Assumption~\ref{missingness_mechanism} implies that data are considered missing at random if the missingness depends only on the observed data; otherwise, they are classified as missing not at random. Assumption~\ref{missingness_mechanism} (i) implies $\mathbb{E}[\Delta Y \mid D=d,R=1,X]=\mathbb{E}[\Delta Y \mid D=d,X]$, so the ATT in Equation~(\ref{full_identification}) can be identified from complete cases after correcting for the response probability. Hence, an inverse probability weighting (IPW) identification formula for the ATT under MAR can be expressed as
\begin{align}
    \tau^{\text{MAR}} := \mathbb{E}\left[\frac{R(D-\Pr(D=1 \mid X))}{\Pr(D=1) \Pr(R=1 \mid D,X)(1-\Pr(D=1 \mid X))}\Delta Y\right]= \tau , \label{mar_iden_basic}
\end{align}
where $\Pr(D=1)$ is the probability of treatment, and $\Pr(R=1 \mid D,X)=\Pr(R=1 \mid \Delta Y, D, X)$ is the response mechanism under MAR, which can be estimated from observed data. Equation~(\ref{mar_iden_basic}) can be decomposed into two components, $R / \Pr(R=1 \mid D,X)$ and $(D-\Pr(D=1 \mid X))\Delta Y / \{\Pr(D=1)(1-\Pr(D=1 \mid X))\}$. The first component is the inverse probability weight for missingness, which handles outcome nonresponse, and the second component is the summand of the IPW identification formula for the ATT as in \parencites{abadie2005semiparametric}{sant2020doubly}. The proof is provided in Appendix~\ref{proof_mar_iden_basic}. It is important to note that the MAR assumption employed in this paper differs from the standard cross-sectional MAR condition $R \indep Y_{1} \mid (D, X)$ because it allows missingness to be associated with $Y_1$ through the fully observed pre-treatment outcome $Y_0$. Identifying the ATT under MNAR is more challenging and requires additional assumptions. Intuitively, when missingness is not at random, $\mathbb{E}[\Delta Y \mid D,R=0,X]$ and $\Pr(R=1 \mid \Delta Y, D,X)$ are needed to identify the ATT, but these two functions cannot be directly identified from observed data without extra information. As one possible solution, we next introduce the ``bespoke'' IV method.

\begin{remark} \label{remark_1}
  \textcite{shin2024difference} employ a ``bespoke'' IV that satisfies three additional conditions. We extend their ``bespoke'' IV conditions by incorporating covariates, as stated in Assumption~\ref{bespoke_assumption} in the Appendix. The central idea of the ``bespoke'' IV method is to express the expectation of outcome evolution among nonrespondents as {\small $\mathbb{E}[\Delta Y \mid D=d,R=0,X]=\mathbb{E}[\Delta Y \mid D=d,R=1,X] + W_d(X)$}, where
{\small 
\begin{equation*}
  W_d(X)= \frac{
\mathbb{E}[\Delta Y \mid D = d, R = 1, \tilde{R} = 1, X]
- \mathbb{E}[\Delta Y \mid D = d, R = 1, \tilde{R} = 0, X]
}
{
\Pr(R = 0 \mid D = d, \tilde{R} = 0, X)
-
\Pr(R = 0 \mid D = d, \tilde{R} = 1, X)
},
\end{equation*}
}
for $d\in\{0,1\}$, where $\tilde{R}$ is a valid binary ``bespoke'' IV. 
From the expression of $W_d(X)$, the numerator captures the difference in expected outcome evolution between respondents with $\tilde{R}=1$ and respondents with $\tilde{R}=0$, while the denominator captures the difference in nonresponse rates across $\tilde{R}$ values. Thus, $W_d(X)$ represents the shift from the respondent mean to the nonrespondent mean within each $(D=d,X)$ stratum. The assumptions required for the covariate-adjusted ``bespoke'' IV DID estimand are provided in Appendix~\ref{DID_bespoke_IV}.
\end{remark}

However, such an additional variable is not always available in practice, particularly in survey datasets with only a limited number of questions. The shadow variable approach makes it possible to estimate the ATT under MNAR without finding an extra auxiliary variable. Assume that the fully observed covariates $X$ can be decomposed as $X=(U,Z)^\top$, such that
\begin{assumption}[ Shadow Variable Property]\label{sv_assumption}
\end{assumption}
\vspace{-8ex}
\begin{align*}
     Z \indep R \mid (\Delta Y, D, U), \quad Z \noindep \Delta Y \mid (D, R, U)
\end{align*}
The vector $U$ denotes the baseline covariates, and $Z$ is termed the shadow variable. The shadow variable assumption follows the framework of \parencites{miao2016varieties}{miao2024identification}, which we extend to the DID design. Intuitively, $Z$ predicts the change in outcome, $\Delta Y$, but has no direct effect on missingness, $R$, after conditioning on $(\Delta Y,D,U)$. This exclusion restriction provides additional information to disentangle the relationship between outcome evolution and missingness, thus enabling identification in MNAR settings. Figure~\ref{fig_causal_diagram} presents causal diagrams that illustrate Assumptions~\ref{missingness_mechanism} and~\ref{sv_assumption}.

{
  \begin{figure}[htbp]
\centering
\begin{tikzpicture}[
  node distance=2cm and 2.5cm,
  every node/.style={draw, circle, minimum size=1cm},
  every path/.style={-{Latex}, thick}
]

  \node (D1) at (0,0) {\(D\)};
  \node (R1) [above right=of D1] {\(R\)};
  \node (DY1) [right=of D1] {\(\Delta Y\)};

  \draw (D1) -- (R1);
  \draw (D1) -- (DY1);

  \node (Z2) at (7,0) {\(Z\)};
  \node (D2) [right=of Z2] {\(D\)};
  \node (DY2) [right=of D2] {\(\Delta Y\)};
  \node (R2) [above right=of D2] {\(R\)};

  \draw (Z2) -- (D2);
  \draw (D2) -- (DY2);
  \draw (D2) -- (R2);
  \draw (DY2) -- (R2);
  \draw[bend right=30] (Z2) to (DY2);

\end{tikzpicture}
\caption{Two causal diagrams illustrating MAR and MNAR (with a valid shadow variable), allowing $X$ or $U$ to have directed arrows to all variables in both diagrams.}
\label{fig_causal_diagram}
\end{figure}
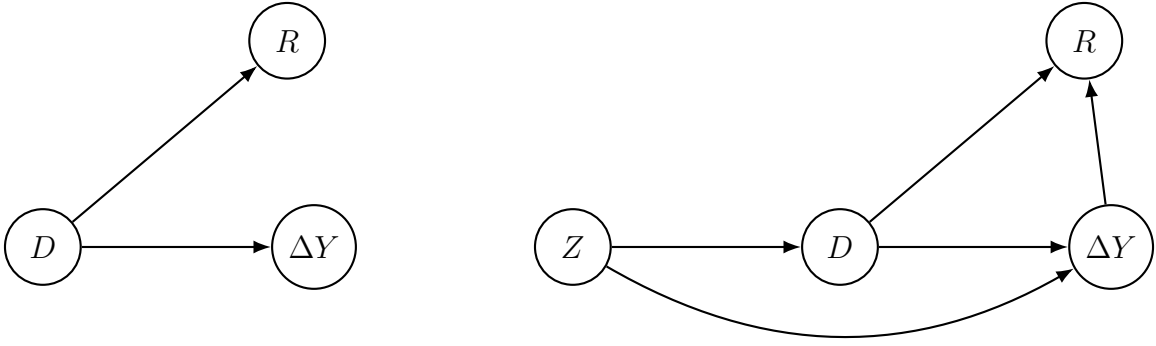
}

It is worth noting that a shadow variable only provides the possibility of identifying the target parameter under the full-data law from the observed-data law; identifiability still needs to be investigated on a case-by-case basis. See the examples in \parencites{wang2014instrumental}{zhao2022versatile} for further details. Here, identifiability means that two different sets of parameters generate two different observed-data distributions, which is necessary for the existence of consistent estimators. Even when the model is identifiable, modeling the missingness mechanism $\Pr(R=1 \mid \Delta Y, D, X)=\Pr(R=1 \mid \Delta Y, D, U)$ based on observed data remains challenging because the mechanism is associated with the potentially unobserved outcome evolution. Consequently, researchers have proposed a variety of methodological approaches targeting different parameters. For instance, \textcite{wang2014instrumental} employ the generalized method of moments (GMM), while \textcite{miao2016varieties} utilize the odds ratio model to estimate the outcome mean. In addition, \textcite{zhao2015semiparametric} propose a pseudo-likelihood method, and \textcite{zhao2022versatile} develop a decorrelated projection technique to estimate regression coefficients. In this paper, following \parencites{miao2016varieties}{miao2024identification}, we use the odds ratio model to identify and estimate the ATT under the DID design. In the next subsection, we check the identifiability of the odds ratio.

\section{Identifiability, Identification and Estimation}
Let $p(\cdot)$ denote a generic probability density function (pdf) or probability mass function (pmf), with the argument and conditioning set made clear from context. The odds ratio is defined as
\begin{equation}
  \OR(\Delta Y, D,U,Z)=\frac{
  p(\Delta Y \mid D , R = 0, U, Z) 
  \cdot 
  p(\Delta Y = 0 \mid D , R = 1, U, Z)
}{
  p(\Delta Y \mid D , R = 1, U, Z) 
  \cdot 
  p(\Delta Y = 0 \mid D , R = 0, U, Z)
}, \label{odds_ratio}
\end{equation}
where $\Delta Y=0$ is a reference value; any other value within the support of $\Delta Y$ can be chosen by the analyst. The odds ratio model encodes the association between the outcome evolution and the missingness process. 
 Compared with the ``bespoke'' IV method, where identification relies on additionally imposed IV restrictions, the odds ratio function involves $p(\Delta Y \mid D=d, R=0, U, Z)$, which is not fully observable from the data. Hence, the identifiability of the odds ratio must be verified first. To achieve identifiability of the odds ratio, we assume:

\begin{assumption}[Identification Conditions] \label{iden_assumption}
\end{assumption}
\vspace{-8ex}
\begin{align*}
  &\mathrm{(i)} \quad 0< \Pr(R=1 \mid \Delta Y,D,U,Z)= \Pr(R=1 \mid \Delta Y,D,U)<1 \ \text{with probability} \ 1 \\
  &\mathrm{(ii)} \quad \text{Completeness of the conditional law} \ p(\Delta Y \mid D,R=1,U,Z) \ \text{with respect to} \ Z
\end{align*}
Assumption~\ref{iden_assumption} (i) implies that $\OR(\Delta Y,D,U,Z)>0$ and $\mathbb{E}\left[\OR(\Delta Y,D,U,Z) \mid D,R=1,U,Z\right] < \infty$ hold with probability 1. Assumption~\ref{iden_assumption} (ii) is the completeness condition: for all square-integrable functions $h(\Delta Y,D,U)$, $\mathbb{E}[h(\Delta Y,D,U) \mid D,R=1,U,Z]=0$ almost surely if and only if $h(\Delta Y,D,U)=0$ almost surely. Assumption~\ref{iden_assumption} (ii) is widely used in the literature for model identifiability; see, e.g., \textcite{newey2003instrumental}, \textcite{d2010new}, and \textcite{zhao2022versatile}. It is satisfied for many common models in which the conditional density of the outcome belongs to an exponential family. Here, we follow the completeness condition of \textcite{miao2024identification}, which involves only observed-data quantities and provides a sufficient condition for nonparametric identifiability of the odds ratio. We derive the following results in the presence of a shadow variable.

\begin{proposition}\label{proposition_1}
If Assumptions~\ref{sv_assumption} and~\ref{iden_assumption} hold, we have
\end{proposition}
\vspace{-8ex}
\begin{align}
  &\OR(\Delta Y,D,U,Z) = \frac{
  \Pr(R = 0 \mid \Delta Y,\, D ,\, U) \cdot 
  \Pr(R = 1 \mid \Delta Y = 0,\, D ,\, U)
}{
  \Pr(R = 1 \mid \Delta Y,\, D ,\, U) \cdot 
  \Pr(R = 0 \mid \Delta Y = 0,\, D ,\, U)
} := \OR(\Delta Y,D,U) \label{OR_representation}\\[1.5ex]
&\mathbb{E}\left[\frac{\OR(\Delta Y,D,U)}{\mathbb{E} \left[ \OR(\Delta Y,D,U) \mid D , R = 1, U \right]} 
 \ \Big| \ D, R = 1, U,Z\right]
=\frac{p(Z \mid D, R = 0, U)}{p(Z \mid D, R = 1, U)},
\label{OR_identifiability}\\[1.5ex]
&\Pr(R = 1 \mid \Delta Y, D, U)
= \left\{\OR(\Delta Y,D,U) \frac{\Pr(R = 0 \mid \Delta Y = 0, D , U)}{\Pr(R = 1 \mid \Delta Y = 0, D, U)}+1\right\}^{-1}
 \label{missingness_mechanism_mnar}\\[1.5ex]
&p(\Delta Y \mid D,R=0,U,Z)=\frac{\OR(\Delta Y,D,U)p(\Delta Y \mid D,R=1,U,Z)}{\mathbb{E}[\OR(\Delta Y,D,U) \mid D,R=1,U,Z]} \label{recover_identification} \\[1.5ex]
&\mathbb{E}\left[\Delta Y \mid D, R=0,U,Z\right]=\frac{\mathbb{E}\left[\OR(\Delta Y,D,U) \Delta Y \mid D, R=1,U,Z\right]}{\mathbb{E}\left[\OR(\Delta Y,D,U) \mid D, R=1,U,Z\right]} \label{expectation_identification}
\end{align}
Equation~(\ref{OR_representation}) shows how the shadow variable enters the odds ratio representation. Equation~(\ref{OR_identifiability}) establishes identifiability of the odds ratio model. In particular, because both $p(Z \mid D=d, R=r, U)$ for $d,\ r\in\{0,1\}$ and $p(\Delta Y \mid D=d, R=1, U, Z)$ are identified from observed data, Equation~(\ref{OR_identifiability}) forms a Type I Fredholm integral equation with a unique solution under the completeness condition. This guarantees nonparametric identification of the odds ratio function; see Appendix~\ref{Appendix_OR_Iden} for details. Equation~(\ref{missingness_mechanism_mnar}) shows how the odds ratio recovers the missingness mechanism under MNAR. Equation~(\ref{recover_identification}) shows that the odds ratio recovers the missing-data distribution from the complete-case distribution, and Equation~(\ref{expectation_identification}) shows that the conditional mean of outcome evolution among incomplete cases is determined by the observed-data distribution and the odds ratio model. The proof of Proposition~\ref{proposition_1} is provided in Appendix~\ref{proof_Proposition1}.

However, a continuous outcome with a binary shadow variable is common in empirical studies and violates the completeness condition. For example, in the household debt study, we use hukou status in 2015 (urban/rural household registration) as the shadow variable. Hukou plausibly satisfies the shadow variable conditions: it predicts household debt changes through differential access to credit markets across registration types, while having no direct effect on a household's decision to respond to the debt question in the survey, conditional on the actual debt change and other covariates. In addition, \textcite{newey2003instrumental} noted that when both $y$ and $z$ are discrete with finite supports $\{y_1, \ldots, y_s\}$ and $\{z_1, \ldots, z_t\}$, respectively, the completeness condition implicitly requires $t \ge s$, that is, the shadow variable must have support no smaller than that of the variable with missing values. Example 1 in \textcite{zhao2022versatile} provides the same result. To address this limitation, and following the examples in \textcite{miao2024identification}, we note that the odds ratio can still be identified without completeness when the odds ratio is modeled parametrically. We illustrate this point with the following example.

\noindent \textbf{Example 1} For simplicity, we consider $X=Z$, and the valid shadow variable $Z$ is binary, $\Delta Y \mid D=d,R=1 \sim \mathcal{N}(\mu_{d},\sigma^2_{d})$ and $\Delta Y \mid D=d,R=1,Z=1 \sim \mathcal{N}(\mu_{dz},\sigma^2_{d})$. Suppose the odds ratio function is specified as $\OR(\Delta Y, D=d;\gamma)= \exp(-\gamma \Delta Y)$. Based on Equation~(\ref{OR_identifiability}), we have:
\begin{align*}
  \exp(-\gamma(\mu_{dz}-\mu_{d})) = \frac{p(Z=1 \mid D=d, R=0)}{p(Z=1 \mid D=d, R=1)},
\end{align*}
where the left-hand side is derived by straightforward calculation and is monotone in $\gamma$ if $Z \noindep \Delta Y \mid D=d, R$ holds ($\mu_{d} \neq \mu_{dz}$), while the right-hand side is a constant identified from the observed data. Hence, the solution for $\gamma$ is unique, which guarantees the identifiability of the odds ratio function. 

\subsection{Identification}
In this subsection, we show how to use the odds ratio to identify the ATT when the response is nonignorable. The identification result is given as follows:

\begin{proposition}\label{proposition_2}
Let Assumptions~\ref{basic_assumption}, \ref{missingness_mechanism} (ii), \ref{sv_assumption}, and~\ref{iden_assumption} hold. The ATT under the MNAR setting can be identified by
\end{proposition}
\vspace{-4ex}
\begin{equation}
\tau^{\text{MNAR}} := \mathbb{E}\left[\frac{R(D-\Pr(D=1 \mid U,Z))}{\Pr(R=1 \mid \Delta Y,D,U)(1-\Pr(D=1 \mid U,Z))}\Delta Y\right]/\Pr(D=1) = \tau  \label{mnar_iden_basic}
\end{equation} 

The proof is provided in Appendix~\ref{proof_Proposition2}. Based on the representation of $\tau^{\text{MNAR}}$, if the response mechanism is MAR and is compatible with the shadow variable restriction, so that $\Pr(R=1 \mid \Delta Y,D,U)=\Pr(R=1 \mid D,U)=\Pr(R=1 \mid D,U,Z)$, then $\tau^{\text{MNAR}}$ reduces to the MAR estimand $\tau^{\text{MAR}}$ with $X=(U,Z)^\top$.

\subsection{Estimation}\label{mnar_estimation_section}
\textbf{Step 1:} Specify the following working models:

$\pi(U,Z;\beta)=\Pr(D=1 \mid U,Z;\beta)$, $\Pr(R=1 \mid \Delta Y=0, D,U;\alpha)$, and $\OR(\Delta Y,D,U;\gamma)$, e.g., logistic models for the propensity score and baseline response mechanism, and $\OR(\Delta Y,D,U;\gamma)=\exp(-\gamma \Delta Y)$.

\noindent \textbf{Step 2:} Construct the missingness mechanism
\begin{equation*}
 q(\Delta Y, D,U;\alpha,\gamma) 
= \left\{\OR(\Delta Y,D,U;\gamma) \frac{\Pr(R = 0 \mid \Delta Y = 0, D , U;\alpha)}{\Pr(R = 1 \mid \Delta Y = 0, D, U;\alpha)}+1\right\}^{-1}
\end{equation*}
 which is used to estimate $\Pr(R = 1 \mid \Delta Y, D,U)$ and is derived from Equation~(\ref{missingness_mechanism_mnar}).

\noindent \textbf{Step 3:} Estimate the parameters $\alpha$, $\beta$ and $\gamma$ by solving the following equations:
\begin{equation*}
  \frac{1}{n}\sum_{i=1}^{n} \left\{\frac{R_i}{q(\Delta Y_i, D_i,U_i;\hat{\alpha},\hat{\gamma})}-1\right\} H(D_i,Z_i,U_i)=0
\end{equation*}
$H(D,Z,U)$ is a user-specified vector of functions with dimension equal to that of $(\alpha,\gamma)$, for example, $H(D,Z,U)=(1,D,Z,U)^\top$. In these equations, terms involving $q(\Delta Y_i,D_i,U_i;\hat{\alpha},\hat{\gamma})$ are evaluated only for respondents with $R_i=1$, for whom $\Delta Y_i$ is observed.
\begin{equation*}
  \frac{1}{n}\sum_{i=1}^{n} \frac{R_i(D_i-\pi(U_i,Z_i;\hat{\beta}))}{q(\Delta Y_i, D_i,U_i;\hat{\alpha},\hat{\gamma})(1-\pi(U_i,Z_i;\hat{\beta}))}G(U_i,Z_i) =0
\end{equation*}
where $\hat{\alpha}$ and $\hat{\gamma}$ are obtained from the previous equation, and $G(U,Z)$ is a user-specified vector of functions with dimension equal to that of $\beta$, for example, $G(U,Z)=(1,Z,U)^\top$.

\noindent \textbf{Step 4:} Based on the identification result in Proposition~\ref{proposition_2}, the DID estimator under MNAR is:
\begin{equation}
  \hat{\tau}^{\text{MNAR}} = n^{-1}\sum_{i=1}^{n} \frac{R_i(D_i-\pi(U_i,Z_i;\hat{\beta}))}{\bar{D}q(\Delta Y_i,D_i,U_i;\hat{\alpha},\hat{\gamma})(1-\pi(U_i,Z_i;\hat{\beta}))}\Delta Y_{i}, \label{mnar_estimator}
\end{equation}
 where $\bar{D}=n^{-1}\sum_{i=1}^n D_i$ estimates $\Pr(D=1)$.

\begin{remark}\label{remark_2}
One may consider constructing an augmented estimator by combining the outcome regression and the IPW representation of the ATT under MNAR, such as
\end{remark}
\vspace{-1cm}
\begin{equation*}
  \mathbb{E}\left[\frac{R(D-\Pr(D=1 \mid U,Z))}{\Pr(R=1 \mid \Delta Y,D,U)(1-\Pr(D=1 \mid U,Z))}(\Delta Y-\mathbb{E}[\Delta Y \mid D=0,U,Z])\right]/\Pr(D=1)
\end{equation*}
By the law of total expectation, $\mathbb{E}[\Delta Y \mid D=0,U,Z]=\sum_{r\in\{0,1\}}\mathbb{E}[\Delta Y \mid D=0,R=r,U,Z]\Pr(R=r \mid D=0,U,Z)$, which can be identified based on Equation~(\ref{expectation_identification}). However, implementing this augmented representation requires additional modeling of the response probability and the outcome regression, which increases the difficulty of correct model specification. Hence, we do not pursue this approach in this paper. The proposed estimator instead has a double-robust-type property with respect to the propensity score and the outcome model, conditional on correct specification of the missingness mechanism, following the idea of \textcite{li2025difference}.

For example, if we specify $G(U,Z)=(1,Z,U)^\top$, the second estimating equation in Step 3 implies that:
  \begin{equation*}
  \frac{1}{n}\sum_{i=1}^{n} \frac{R_i(D_i-\pi(U_i,Z_i;\hat{\beta}))}{q(\Delta Y_i, D_i,U_i;\hat{\alpha},\hat{\gamma})(1-\pi(U_i,Z_i;\hat{\beta}))} (1,Z_i,U_i)^\top \theta =0,
\end{equation*}
where $\theta$ can take any value including the true parameter $\theta^*$. Then, the proposed estimator can be rewritten as: 
\begin{equation}
  \hat{\tau}^{\text{MNAR}} = n^{-1}\sum_{i=1}^{n} \frac{R_i(D_i-\pi(U_i,Z_i;\hat{\beta}))}{\bar{D}q(\Delta Y_i,D_i,U_i;\hat{\alpha},\hat{\gamma})(1-\pi(U_i,Z_i;\hat{\beta}))}(\Delta Y_{i}-(1,Z_i,U_i)^\top \theta), \label{mnar_estimator_cbps}
\end{equation}
If the outcome model is linear in covariates, i.e., $\mathbb{E}[\Delta Y \mid D=0,U,Z]=(1,Z,U)^\top \theta^*$, then $\hat{\tau}^{\text{MNAR}}$ in Equation~(\ref{mnar_estimator_cbps}) is consistent even if the propensity score $\pi(U,Z;\beta)$ is misspecified. Hence, the final estimator is doubly robust: it is consistent if both the missingness mechanism $q(\Delta Y, D,U;\alpha,\gamma)$ and the propensity score $\pi(U,Z;\beta)$ are correctly specified; or if the outcome model is linear in covariates and the missingness mechanism $q(\Delta Y, D,U;\alpha,\gamma)$ is correctly specified. Next, we shift our focus to establishing the asymptotic normality of the proposed estimator.

\begin{theorem}\label{theorem_1}
Suppose that all working models are correctly specified, i.e., $q(\Delta Y,D,U;\alpha^*,\gamma^*)=\Pr(R=1 \mid \Delta Y,D,U)$ and $\pi(U,Z;\beta^*)= \Pr(D=1 \mid U,Z)$, and that the assumptions in Proposition~\ref{proposition_2} hold. Additionally, assume the stacked estimating equations are locally identified, the relevant derivative matrix is nonsingular, and the regularity conditions stated in Assumptions A.1 and A.2 of \textcite{sant2020doubly} are satisfied. Then
\end{theorem}
\vspace{-4ex}
\begin{equation}
  \sqrt{n}(\hat{\tau}^{\text{MNAR}}-\tau) \xrightarrow{d} \mathcal{N}(0,V_{\tau}),  
\end{equation}
where $V_{\tau}$ is the asymptotic variance of $\hat{\tau}^{\text{MNAR}}$, obtained from the sandwich variance in Appendix~\ref{proof_theorem1}. Theorem~\ref{theorem_1} shows that the proposed estimator is consistent and asymptotically normal when all working models are correctly specified. The regularity conditions primarily require that the nuisance parameters satisfy asymptotic linearity, that the estimating functions have finite second moments, and that the derivative matrix is nonsingular, all of which are standard in the M-estimation literature.

\begin{proposition}\label{proposition5_MNAR}
Suppose the response mechanism is MAR and remains compatible with the shadow variable exclusion restriction, so that
$\Pr(R=1 \mid \Delta Y,D,U,Z)=\Pr(R=1 \mid D,U,Z)$ and 
$\Pr(R=1 \mid \Delta Y,D,U,Z)=\Pr(R=1 \mid \Delta Y,D,U)$.
Then the response mechanism also satisfies $\Pr(R=1 \mid D,U,Z)=\Pr(R=1 \mid D,U)$. Consequently, $\Pr(R=1 \mid \Delta Y,D,U)=\Pr(R=1 \mid D,U)$, and the MNAR estimand in Equation~(\ref{mnar_iden_basic}) reduces to the MAR IPW estimand with $X=(U,Z)$ under this compatible MAR submodel. Thus, $\tau^{\text{MNAR}}=\tau^{\text{MAR}}=\tau$.

For the modeling procedure, the missingness model $\Pr(R=1 \mid \Delta Y, D, U; \alpha,\gamma)$ remains correctly specified under this MAR submodel if the odds ratio model nests MAR as a special case. That is, there exists a value $\gamma^*$ such that $\OR(\Delta Y,D,U;\gamma^*)=1$. For instance, if $\OR(\Delta Y,D,U;\gamma)=\exp(-\gamma\Delta Y)$, then $\gamma^*=0$ yields $\OR=1$. Under this condition, Equation~(\ref{missingness_mechanism_mnar}) implies $\Pr(R=1 \mid \Delta Y,D,U;\alpha^*,\gamma^*)=\Pr(R=1 \mid D,U;\alpha^*)$. Therefore, the proposed MNAR estimator remains consistent when the true missingness mechanism lies in this MAR submodel.
\end{proposition}

\section{Simulation Study}

We conduct Monte Carlo simulations to evaluate the finite-sample performance of the proposed MNAR estimator and to compare it against two benchmark estimators: the naive DID estimator and the MAR DID estimator based on Equation~(\ref{mar_iden_basic}). Two scenarios are considered, $\gamma=0$ (MAR) and $\gamma=-0.3$ (MNAR). For each scenario we run $10{,}000$ Monte Carlo replications, and in every replication an independent sample of size $n=10{,}000$ is drawn afresh from the data-generating process described below.

\subsection{Data Generating Process}

We generate baseline covariates $U_1\sim\mathcal{N}(0,1)$ and $U_2\sim\mathrm{Bernoulli}(0.5)$, together with a binary shadow variable $Z\sim\mathrm{Bernoulli}(\mathrm{expit}(0.3U_1+0.2U_2))$. Treatment is assigned via a logistic propensity score that includes the shadow variable, $D\sim\mathrm{Bernoulli}(\mathrm{expit}(-0.2+0.5U_1+0.2U_2+0.2Z))$, yielding a treatment rate of approximately 50\%. The pre-treatment outcome is $Y_{0}=10+2U_1+3U_2+\varepsilon_0$ with $\varepsilon_0\sim\mathcal{N}(0,1)$. The potential outcomes are $Y_{1}(0)=Y_{0}+5Z+\varepsilon_1$ and $Y_{1}(1)=Y_{1}(0)+\tau$, where $\varepsilon_1\sim\mathcal{N}(0,1)$ and the true ATT is $\tau=3$.

The missingness mechanism is generated as follows. Let $p_0=\mathrm{expit}(2.1+0.3D+0.3U_1+0.4U_2)$ denote the baseline response probability $\Pr(R=1\mid\Delta Y=0,D,U)$, and let the odds ratio model be $\OR(\Delta Y;\gamma)=\exp(-\gamma\Delta Y)$, with $\gamma=0$ or $\gamma=-0.3$. The actual response probability is then $\Pr(R=1\mid\Delta Y,D,U)=\bigl\{\exp(-\gamma\Delta Y)(1-p_0)/p_0+1\bigr\}^{-1}$, with $R\sim\mathrm{Bernoulli}(\Pr(R=1\mid\Delta Y,D,U))$. Hence $Z$ is independent of $R$ given $(\Delta Y,D,U)$ but associated with $\Delta Y$ among respondents conditional on $(D,R,U)$, as required by Assumption~\ref{sv_assumption}. By construction, $\gamma=0$ recovers the MAR case ($\OR=1$, missingness independent of $\Delta Y$ given $(D,U)$), whereas $\gamma=-0.3$ introduces MNAR dependence.
We consider three estimators in each scenario:
\begin{enumerate}[label=(\roman*)]
  \item The complete-case naive DID estimator ($\hat{\tau}^{\text{DID}}$), computed as the difference in mean $\Delta Y$ between observed treated ($R=1, D=1$) and observed control ($R=1, D=0$) units, without any covariate adjustment.
  \item The MAR estimator ($\hat{\tau}^{\text{MAR}}$). It models the propensity score $\Pr(D=1\mid U,Z)$ and the response probability $\Pr(R=1\mid D,U,Z)$ separately by logistic regression, and plugs them into the identification formula $\tau^{\text{MAR}}$ in Equation~(\ref{mar_iden_basic}).
  \item The proposed MNAR DID estimator ($\hat{\tau}^{\text{MNAR}}$), which follows the four-step procedure outlined in Subsection~\ref{mnar_estimation_section}.
\end{enumerate}

\subsection{Results and Discussion}

\begin{figure}[htbp]
  \centering
  \includegraphics[width=0.6\textwidth]{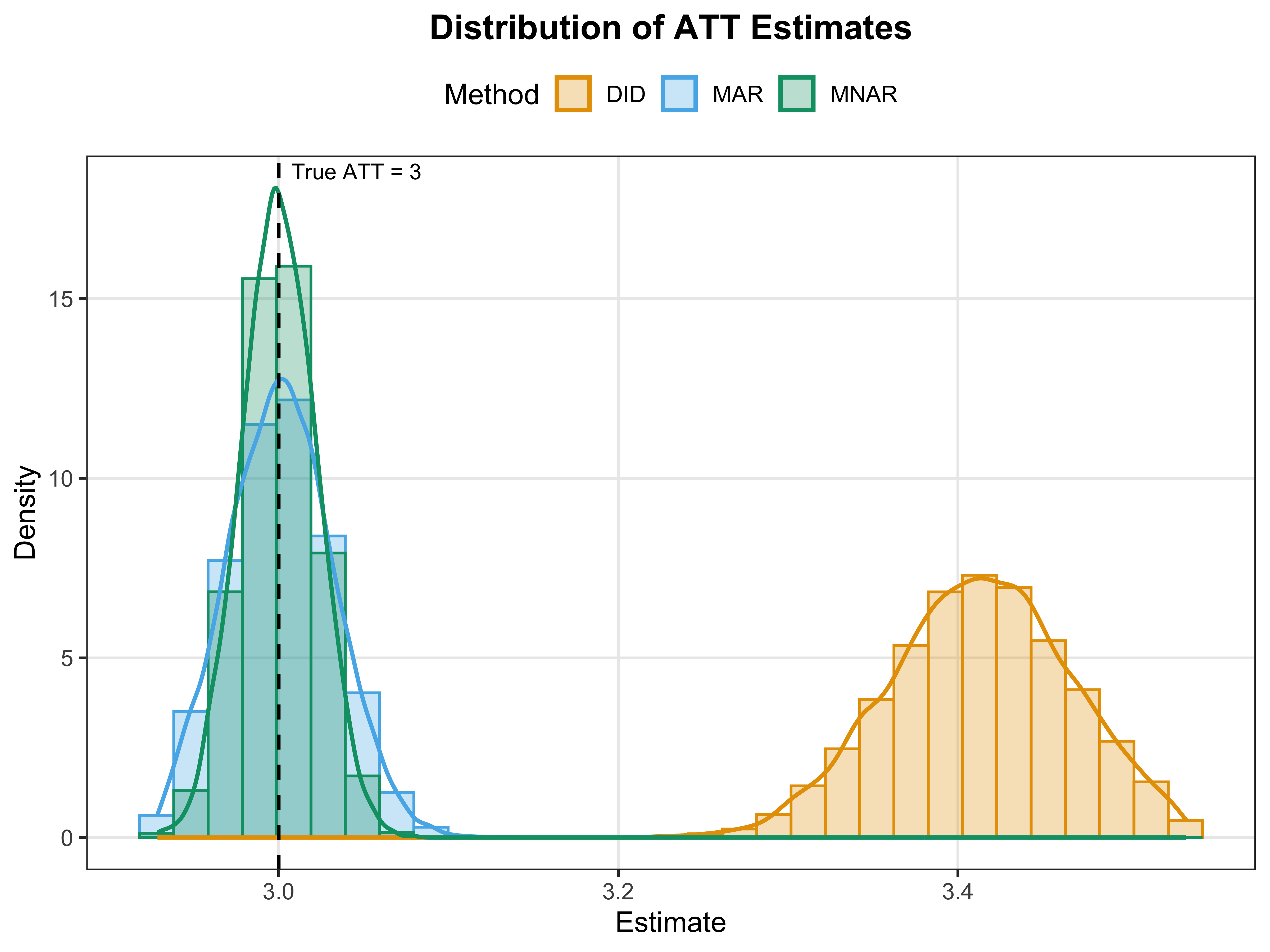}
  \caption{Sampling distributions of ATT estimators under $\gamma=0$ (MAR). Based on $10{,}000$ Monte Carlo replications with $n=10{,}000$ per replication. The dashed line marks the true ATT $=3$.}
  \label{fig:sim_gamma0}
\end{figure}

\begin{figure}[htbp]
  \centering
  \includegraphics[width=0.6\textwidth]{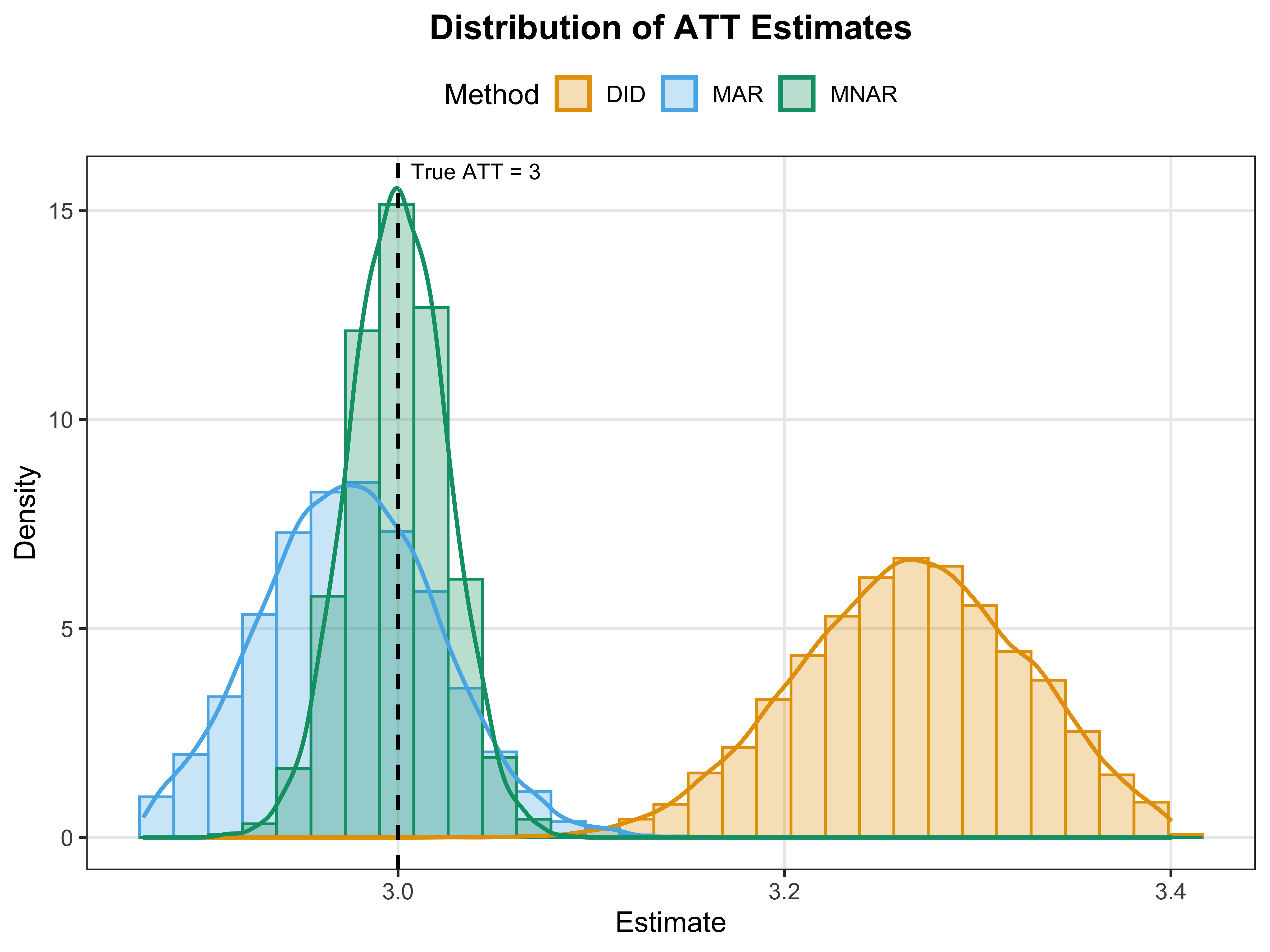}
  \caption{Sampling distributions of ATT estimators under $\gamma=-0.3$ (MNAR). Based on $10{,}000$ Monte Carlo replications with $n=10{,}000$ per replication. The dashed line marks the true ATT $=3$.}
  \label{fig:sim_gamma03}
\end{figure}

Figures~\ref{fig:sim_gamma0} and~\ref{fig:sim_gamma03} display the sampling distributions of the three estimators under $\gamma=0$ and $\gamma=-0.3$, respectively. Several observations follow.

Across both scenarios, the naive DID estimator is substantially biased, with its sampling distribution centered near $3.4$ under $\gamma=0$ and near $3.3$ under $\gamma=-0.3$. The source of this bias is that the naive DID corrects for neither propensity score imbalance nor differential missingness in $\Delta Y$, thereby inheriting the resulting bias. The behavior of the MAR and MNAR estimators under these two scenarios aligns directly with the theoretical results presented in the previous section.
Under $\gamma=0$ (Figure~\ref{fig:sim_gamma0}), the odds ratio $\exp(0 \cdot \Delta Y)$ collapses to one, so the MNAR estimand $\tau^{\text{MNAR}}$ reduces to $\tau^{\text{MAR}}$, as established in Proposition~\ref{proposition5_MNAR}. The simulation reproduces this nesting in finite samples: both estimators are centered at the true ATT $\tau=3$, confirming that adopting the more flexible MNAR procedure introduces no bias when the compatible MAR model holds.
Under $\gamma=-0.3$ (Figure~\ref{fig:sim_gamma03}), by additionally modeling the odds ratio through the GMM step, $\hat{\tau}^{\text{MNAR}}$ removes the residual selection bias and produces a sampling distribution more closely centered at $\tau=3$ than $\hat{\tau}^{\text{MAR}}$, consistent with the consistency guarantee of Theorem~\ref{theorem_1}.

In addition to recovering the location accurately, the MNAR estimator exhibits desirable variability properties in these simulations. In both scenarios, its sampling distribution is approximately symmetric and bell-shaped around the true ATT, consistent with the asymptotic normality result established in Theorem~\ref{theorem_1}. The MNAR distribution is no more dispersed than the MAR distribution in the displayed simulations and is noticeably more concentrated under $\gamma=-0.3$, where it achieves both bias correction and variance reduction. This finite-sample efficiency gain may be attributed to the calibrated propensity score from Step~3 of the four-step procedure, which enforces exact moment balance for $(1,Z,U)$ in the weighted sample and has been shown to perform well in finite samples \parencite{li2025difference}.

\section{Empirical Example}
	\label{sec:empirical}

	We draw on \textcite{deng2021does}, who study the effect of China's two-child policy relaxation on household debt using a propensity score matching (PSM) DID design with CHFS panel data (2015 and 2017 waves). Their central finding is that a household birth event does not significantly increase debt under a standard DID specification, but the effect becomes statistically significant after PSM, suggesting that selection into treatment confounds the naive comparison.

  There are two concerns with interpreting these estimates at face value. First, missingness in the debt outcome is nontrivial and unlikely to be random.  
  Households experiencing a deterioration in their financial position may be systematically less willing to disclose debt information in subsequent survey rounds, rendering the MAR assumption implausible. Second, the treatment group is inherently small: treated households, namely those reporting a birth event after the policy relaxation, constitute approximately $2.1\%$ of the panel sample. This sparsity is further compounded by PSM, which discards a substantial portion of the control group to achieve covariate balance, placing additional strain on the precision of the estimates and raising concerns about the reliability of inference in the matched sample.

  The MNAR DID approach proposed in this paper alleviates both concerns simultaneously. By modeling the missingness mechanism, it corrects for MNAR without discarding nonresponse observations, and by replacing PSM with internal reweighting via propensity score, it retains the full sample throughout estimation. Following \textcite{deng2021does}, we employ the 2015 and 2017 waves of the CHFS and use the same set of variables that they chose. After restricting to households with valid outcome and covariate information, the estimation sample comprises 7,917 observations, of which $8.2\%$ have missing debt outcomes and approximately $2.1\%$ belong to the treatment group. Table~\ref{tab:setup} summarizes the variable definitions and the shadow variable diagnostics, while Table~\ref{tab:svdid} reports the benchmark estimates from \textcite{deng2021does}, alongside the results of the MAR DID estimator and our proposed MNAR DID estimator.

\begin{table}[htbp]
\centering
\caption{Variable Definitions and Shadow Variable Diagnostics}
\label{tab:setup}
\small
\setlength{\tabcolsep}{10pt}
\renewcommand{\arraystretch}{1.25}
\begin{threeparttable}
\begin{tabular*}{0.98\textwidth}{@{\extracolsep{\fill}} l l@{}}
\toprule
\textbf{Component} & \textbf{Definition / value} \\
\midrule
\multicolumn{2}{@{}l}{\textit{Variables}} \\[2pt]
\quad Outcome ($Y$)         & $\log(\text{household debt}+1)$, observed in 2015 (pre) and 2017 (post) \\
\quad Treatment ($D$)       & Birth event after the two-child policy ($=1$ if a newborn after 2016) \\
\quad Missingness ($R$)     & $=1$ if the 2017 debt outcome is reported, $=0$ otherwise \\
\quad Shadow variable ($Z$) & Hukou (household registration) status, 2015 \\
\quad Covariates ($U$)      & Age, Education, Health, Married, Household Income and Size \\
\addlinespace[4pt]
\multicolumn{2}{@{}l}{\textit{Sample and Shadow Variable Diagnostics}} \\[2pt]
\quad Estimation sample size                   & $7{,}917$ \\
\quad Missing rate              & $8.2\%$ \\
\quad Treatment rate             & $2.1\%$ \\
\quad Corr$(Z,\Delta Y)$ among respondents     & $-0.037$ \\
\quad MNAR intensity parameter $\hat{\gamma}$  & $-0.458$ \\
\bottomrule
\end{tabular*}
\begin{tablenotes}
\small
\item \textit{Notes}: All variables are constructed from the 2015 and 2017 waves of the CHFS,
      following \textcite{deng2021does}. The reported correlation is computed on the observed
      subsample ($R=1$) only. All covariates are measured in the 2015 wave.
\end{tablenotes}
\end{threeparttable}
\end{table}

  As summarized in Table~\ref{tab:setup}, we use household's hukou (household registration) status in 2015 as the shadow variable $Z$. This choice reflects a genuine data constraint: among the covariates available in the CHFS, hukou satisfies the shadow variable conditions most credibly. Its validity rests on two conditions. First, hukou must be associated with the outcome change $\Delta Y$: urban-registered households typically enjoy broader access to formal credit markets, lower borrowing costs, and stronger collateral backing, so hukou systematically predicts how household debt evolves over time. Second, hukou must be independent of the missingness mechanism conditional on $\Delta Y$, $D$, and $U$: once the actual debt change and household characteristics are accounted for, whether a household reports its debt is governed by its financial situation rather than by its registration status per se.

  Among respondents, the sample correlation between hukou and $\Delta Y$ is only $-0.037$. We do not treat this as evidence of a violation of the shadow variable condition. The respondent-level correlation is computed on a selected subsample ($R = 1$) and may be attenuated by nonignorable selection. More importantly, identification under our parametric specification $\mathrm{OR}(\Delta Y;\,\gamma) = \exp(-\gamma\Delta Y)$ does not require a strong marginal correlation: as shown in Example~1, a unique solution for $\gamma$ exists whenever hukou shifts the conditional mean of $\Delta Y$ among respondents by any nonzero amount, which the data support. The cost of a weak shadow variable is reduced precision rather than identification failure, as reflected in the wider bootstrap standard error for the MNAR estimator (0.454 vs.\ 0.384 for MAR). Consistent with this, the estimated intensity $\hat{\gamma} = -0.458$ is economically meaningful: households experiencing larger debt increases are systematically more likely to leave their outcomes unreported, so that a MAR-based estimator would be biased upward.

  Table~\ref{tab:svdid} reports four estimators: the canonical DID, the PSM DID of \textcite{deng2021does}, the MAR DID, and our proposed MNAR DID estimator. The canonical and PSM estimates differ markedly, equal to $0.269$ and $1.632$, respectively. Because treated households account for only $2.1\%$ of the sample and matching discards most of the unmatched control units, the PSM estimate is identified from a thin matched subsample and should be interpreted with caution. We therefore concentrate on the comparison between the MAR and MNAR estimates, which equal $0.584$ and $0.234$.

  The estimated missingness intensity $\hat{\gamma} = -0.458$ departs substantially from zero, implying that households with larger debt increases are more likely to leave their debt outcome unreported. Under such nonignorable selection, the MAR estimator is biased upward. By reweighting the observed records by $1/\hat{q}$ prior to forming the difference-in-differences contrast, the MNAR estimator corrects for this selection and lowers the estimated effect from $0.584$ to $0.234$, suggesting that outcome-dependent nonresponse is a quantitatively important source of bias in this setting.

  The bootstrap standard errors are $0.384$ for the MAR estimate and $0.454$ for the MNAR estimate, and the corresponding $95\%$ percentile bootstrap confidence intervals, $[-0.175,\ 1.339]$ and $[-0.524,\ 1.282]$ over $10{,}000$ replications, are wide and contain zero.\footnote{The wide intervals reflect two compounding sources of imprecision: the weak shadow variable, which limits the information available for GMM identification of $\gamma$, and the small number of observed treated units (approximately 153 out of $7{,}917$ observations, given a $2.1\%$ treatment rate and $91.8\%$ response rate).} We accordingly refrain from drawing conclusions about statistical significance, though the direction of the MNAR correction is consistent with the estimated pattern of selection.
 
  \begin{table}[htbp]
\centering
\caption{MNAR DID Estimation Results and Benchmark Comparison}
\label{tab:svdid}
\small
\setlength{\tabcolsep}{8pt}
\renewcommand{\arraystretch}{1.3}
\begin{threeparttable}
\begin{tabular*}{0.92\textwidth}{@{\extracolsep{\fill}}l c c c@{}}
\toprule
 & \textbf{Estimate} & \textbf{SE} & \textbf{95\% CI} \\
\midrule
\multicolumn{4}{@{}l}{\textit{Benchmark estimator}} \\[2pt]
\quad Canonical DID & $0.269$ & $(0.256)$ & $[-0.233,\ 0.771]$ \\
\quad PSM DID       & $1.632$ & $(0.410)$ & $[0.828,\ 2.436]$ \\
\addlinespace[5pt]
\multicolumn{4}{@{}l}{\textit{Proposed estimator}} \\[2pt]
\quad MAR DID  & $0.584$ & $(0.384)$ & $[-0.175,\ 1.339]$ \\
\quad MNAR DID & $0.234$ & $(0.454)$ & $[-0.524,\ 1.282]$ \\
\bottomrule
\end{tabular*}
\begin{tablenotes}
\small
\item \textit{Notes}: The outcome variable is $\log(\text{debt}+1)$. The benchmark estimates replicate those reported in \textcite{deng2021does}, Table~4. For the proposed estimators, the figures in parentheses are bootstrap standard errors and the $95\%$ confidence intervals are percentile bootstrap intervals, both based on $10{,}000$ replications. 
\end{tablenotes}
\end{threeparttable}
\end{table}

  Our MNAR DID results offer several advantages over the benchmark. First, beyond point estimation, the shadow variable-based DID framework directly provides information on the missingness mechanism itself: the estimated $\hat{\gamma} = -0.458$ indicates a nonnegligible MNAR pattern, and the gap between our MAR ($0.584$) and MNAR ($0.234$) estimates shows that ignoring MNAR can considerably distort the treatment effect estimate. Second, the PSM-DID approach in \textcite{deng2021does} suffers from potentially severe sample loss due to the small treatment group (approximately $2.1\%$ of the sample), as matching discards unmatched control units and may further reduce the effective sample. Our MNAR DID retains the full sample by replacing matching with internal reweighting via $\hat{q}$ and $\hat{\pi}$, preserving all available information. Third, our MNAR estimate of $0.234$ sits below both the canonical DID ($0.269$) and the PSM-DID ($1.632$) of \textcite{deng2021does}, consistent with the view that correcting for outcome-dependent nonresponse attenuates rather than amplifies the estimated policy effect.

\section{Conclusion}

This paper develops a semiparametric DID framework for estimating the ATT when post-treatment outcomes are subject to nonignorable missingness. Building on the shadow variable approach, we show that a fully observed variable that predicts outcome evolution but has no direct effect on response, conditional on the outcome evolution and baseline covariates, can be used to identify the missingness mechanism and recover the ATT under MNAR. We establish identification through an odds ratio representation, propose a feasible GMM-based estimation procedure, and show that the resulting estimator is consistent and asymptotically normal under regularity conditions. 

The proposed framework complements existing DID methods for incomplete panel outcomes. Under a compatible MAR submodel, the MNAR estimand reduces to the familiar MAR IPW DID estimand, so the method nests the ignorable missingness case. When the outcome is truly MNAR, however, the shadow variable provides information that cannot be recovered from complete cases alone. The simulation study confirms this behavior: the proposed estimator remains centered around the true ATT under MAR and substantially reduces selection bias under MNAR. The empirical application to the effect of China's two-child policy on household debt further illustrates the practical relevance of the method. In that setting, the estimated missingness mechanism suggests nonnegligible outcome-dependent nonresponse, and the MNAR correction yields a more moderate treatment effect than the MAR benchmark while retaining the full estimation sample.

Several extensions remain for future work. A first and important direction is to derive the efficient influence function for the MNAR DID parameter, paralleling the semiparametric efficiency theory for nonignorable missing data with a shadow variable in \textcite{miao2024identification}. Such a result would characterize the efficiency bound for the ATT in the DID setting, clarify whether the proposed estimator attains that bound, and guide the construction of locally efficient or multiply robust estimators that combine the odds ratio, response model, propensity score, and outcome regression. This extension would also provide a principled basis for using flexible machine learning methods with cross-fitting while preserving valid inference; see, e.g., \textcite{chang2020double}. Other useful directions include developing sensitivity analysis for possible violations of the shadow variable restriction, extending the framework to multi-period and multi-treatment DID designs, and allowing missingness in both pre- and post-treatment outcomes.
\clearpage
\printbibliography
	\clearpage

\section*{Appendix}
\appendix
\refstepcounter{section}
\subsection{Proof of MAR ATT Identification}
\label{proof_mar_iden_basic}
Based on Assumptions~\ref{basic_assumption} and~\ref{missingness_mechanism} (i), we have
\begin{align*}
\tau^{\text{MAR}}=&\mathbb{E}\left[\frac{R(D-\Pr(D=1 \mid X))}{\Pr(D=1) \Pr(R=1 \mid D,X)(1-\Pr(D=1 \mid X))}\Delta Y\right]\\
=&\mathbb{E}\left[\frac{R(D-\Pr(D=1 \mid X))}{\Pr(D=1) \Pr(R=1 \mid \Delta Y, D,X)(1-\Pr(D=1 \mid X))}\Delta Y\right]\\
=&\mathbb{E}\left[\frac{(D-\Pr(D=1 \mid X))}{\Pr(D=1)(1-\Pr(D=1 \mid X))}\Delta Y\right]\\
=&\mathbb{E}\left[\frac{D}{\Pr(D=1)}\Delta Y\right]-\mathbb{E}\left[\frac{(1-D)\Pr(D=1 \mid X)}{\Pr(D=1)(1-\Pr(D=1 \mid X))}\Delta Y \right]\\
=&\mathbb{E}\left[Y_1(1)-Y_0 \mid D=1\right]-\mathbb{E}\left[\mathbb{E}[Y_1(0)-Y_0\mid D=0,X]\mid D=1\right]\\
=&\mathbb{E}\left[Y_1(1)-Y_0 \mid D=1\right]-\mathbb{E}\left[\mathbb{E}[Y_1(0)-Y_0\mid D=1,X]\mid D=1\right]\\
=&\tau
\end{align*}

\subsection{DID estimand with a bespoke instrumental variable}
\label{DID_bespoke_IV}
Following \textcite{shin2024difference}, we assume that the binary indicator $\tilde{R}$ is available and satisfies the following conditions:
\begin{assumptionA}[Bespoke IV Assumption] \label{bespoke_assumption}
\end{assumptionA}
\vspace{-8ex}
{\footnotesize

\begin{align*}
    &\mathrm{(i)} \quad \Pr(R = 0 \mid D = d, X, \tilde{R} = 0) 
\neq \Pr(R = 0 \mid D = d, X, \tilde{R} = 1) \quad & \text{(Relevance to missingness)}\\
 &\mathrm{(ii)} \quad \mathbb{E}[\Delta Y \mid D = d, X,\tilde{R} = 0] 
= \mathbb{E}[\Delta Y \mid D = d, X,\tilde{R} = 1] \quad & \text{(Independence from outcome evolution)}\\
  &  \mathrm{(iii)} \quad 
  \begin{aligned}[t]
\mathbb{E}[\Delta Y \mid D = d, X, \tilde{R} = 0, R = 1] 
& - \mathbb{E}[\Delta Y \mid D = d, X, \tilde{R} = 0, R = 0] \\
= 
\mathbb{E}[\Delta Y \mid D = d, X, \tilde{R} = 1, R = 1] 
& - \mathbb{E}[\Delta Y \mid D = d, X, \tilde{R} = 1, R = 0] 
\end{aligned}
 \quad & \text{(Bias homogeneity)}
\label{identification_assumptions}
\end{align*}

}
For $d\in\{0,1\}$, define

\begin{align*}
W_d(X)=
\frac{
\mathbb{E}[\Delta Y \mid D=d, X,\tilde{R}=1,R=1]
-\mathbb{E}[\Delta Y \mid D=d, X,\tilde{R}=0,R=1]
}{
\Pr(R=0 \mid D=d,X,\tilde{R}=0)
-\Pr(R=0 \mid D=d,X,\tilde{R}=1)
}.
\end{align*}

Based on Assumptions~\ref{basic_assumption} and~\ref{bespoke_assumption}, the ATT can be identified as

\begin{align*}
\text{ATT}
=&\ \mathbb{E}_X\Big[
\mathbb{E}[\Delta Y \mid D=1,X,R=1]
+ W_1(X)\Pr(R=0 \mid D=1,X)
\mid D=1\Big] \\
&-\mathbb{E}_X\Big[
\mathbb{E}[\Delta Y \mid D=0,X,R=1]
+ W_0(X)\Pr(R=0 \mid D=0,X)
\mid D=1\Big].
\end{align*}

 The proof follows by applying an argument similar to that of Theorem 1 in \textcite{shin2024difference}.

\subsection{Proof of Proposition~\ref{proposition_2}}
\label{proof_Proposition2}
Let $q(\Delta Y,D,U)=\Pr(R=1\mid \Delta Y,D,U)$. Because $q$ does not depend on $Z$ under Assumption~\ref{sv_assumption}, iterated expectation gives

\begin{align*}
\mathbb{E}\left[
\frac{R(D-\Pr(D=1\mid U,Z))}{\Pr(D=1) q(\Delta Y,D,U)(1-\Pr(D=1\mid U,Z))}\Delta Y
\right]
=
\mathbb{E}\left[
\frac{D-\Pr(D=1\mid U,Z)}{\Pr(D=1)(1-\Pr(D=1\mid U,Z))}\Delta Y
\right]= \tau.
\end{align*}

 The last equality follows from the same argument as in Appendix~\ref{proof_mar_iden_basic}, with $X=(U,Z)^\top$.

\subsection{Proof of Proposition~\ref{proposition_1}} 
\label{proof_Proposition1}
Let Assumptions~\ref{sv_assumption} and~\ref{iden_assumption} hold. Then
\begin{align*}
\OR(\Delta Y,D, U, Z) 
&= \frac{ 
    p(\Delta Y \mid D, R = 0, U, Z) \cdot p(\Delta Y = 0 \mid D, R = 1, U, Z)
}{
    p(\Delta Y \mid D, R = 1, U, Z) \cdot p(\Delta Y = 0 \mid D, R = 0, U, Z)
} \\[1ex]
&= \frac{ 
    \frac{\Pr(R = 0 \mid \Delta Y, D, U, Z) \cdot p(\Delta Y \mid D, U, Z)}{\Pr(R = 0 \mid D, U, Z)}
    \cdot 
    \frac{\Pr(R = 1 \mid \Delta Y = 0, D, U, Z) \cdot p(\Delta Y = 0 \mid D, U, Z)}{\Pr(R = 1 \mid D, U, Z)}
}{
    \frac{\Pr(R = 1 \mid \Delta Y, D, U, Z) \cdot p(\Delta Y \mid D, U, Z)}{\Pr(R = 1 \mid D, U, Z)}
    \cdot 
    \frac{\Pr(R = 0 \mid \Delta Y = 0, D, U, Z) \cdot p(\Delta Y = 0 \mid D, U, Z)}{\Pr(R = 0 \mid D, U, Z)}
} \\[1ex]
&= \frac{
    \Pr(R = 0 \mid \Delta Y, D, U) \cdot \Pr(R = 1 \mid \Delta Y = 0, D, U)
}{
    \Pr(R = 1 \mid \Delta Y, D, U) \cdot \Pr(R = 0 \mid \Delta Y = 0, D, U)
} \\[1ex]
&:= \OR(\Delta Y,D,U)
\end{align*}
 Proof of Equation~(\ref{OR_identifiability}).
\begin{footnotesize}
\begin{align*}
&\mathbb{E} \left[ 
\frac{\OR(\Delta Y,D,U)}{
\mathbb{E} \left[ \OR(\Delta Y,D,U) \mid D , R = 1, U \right]} 
\;\middle|\; D, R = 1, U, Z 
\right] \\
=& \mathbb{E} \left[
\frac{
\Pr(R = 0 \mid \Delta Y, D, U) / \Pr(R = 1 \mid \Delta Y, D, U)
}{
\mathbb{E}\left[\Pr(R = 0 \mid \Delta Y, D, U) / \Pr(R = 1 \mid \Delta Y, D, U) \mid D, R=1, U\right]
} 
\;\middle|\; D, R = 1, U, Z
\right] \\
=& \int 
\frac{ 
\Pr(R = 0 \mid \Delta Y, D, U) 
}{ 
\Pr(R = 1 \mid \Delta Y, D, U)
} 
f(\Delta Y \mid D, R = 1, U, Z) \, d\Delta Y 
\bigg/ 
\int 
\frac{ 
\Pr(R = 0 \mid \Delta Y, D, U) 
}{ 
\Pr(R = 1 \mid \Delta Y, D, U)
} 
f(\Delta Y \mid D, R = 1, U) \, d\Delta Y \\[1ex]
=& 
\frac{
\int 
 \Pr(R = 0 \mid \Delta Y, D, U) 
f(\Delta Y \mid D, U, Z) \, d\Delta Y
}{
\Pr(R = 1 \mid D, U, Z)
}
\bigg/ 
\frac{
\int 
\Pr(R = 0 \mid \Delta Y, D, U)
f(\Delta Y \mid D, U) \, d\Delta Y
}{
\Pr(R = 1 \mid D, U)
} \\[1ex]
=& 
\frac{
\Pr(R = 0 \mid D, U, Z) / \Pr(R = 1 \mid D, U, Z)
}{
\Pr(R = 0 \mid D, U) / \Pr(R = 1 \mid D, U)
} 
= 
\frac{ 
p(Z \mid D, R = 0, U) 
}{
p(Z \mid D, R = 1, U)
}
\end{align*}
\end{footnotesize}

The proof of Equation~(\ref{missingness_mechanism_mnar}) follows directly from Equation~(\ref{OR_representation}).

 Proof of Equation~(\ref{recover_identification}).
\begin{align*}
&\frac{
  \OR(\Delta Y,D,U) p(\Delta Y \mid D, R = 1, U, Z)
}{
  \mathbb{E} \left[ \OR(\Delta Y,D,U) \mid D, R = 1, U, Z \right]
} \\[1ex]
=& \frac{
   \Pr(R = 0 \mid \Delta Y, D, U) / \Pr(R = 1 \mid \Delta Y, D, U) 
   p(\Delta Y \mid D, R = 1, U, Z)
}{
  \int 
     \Pr(R = 0 \mid \Delta Y, D, U) / \Pr(R = 1 \mid \Delta Y, D, U) 
     p(\Delta Y \mid D, R = 1, U, Z) \, d\Delta Y
} \\[1ex]
=& \frac{
  \Pr(R = 0 \mid \Delta Y, D, U) p(\Delta Y \mid D, U, Z)
}{
  \int \Pr(R = 0 \mid \Delta Y, D, U) p(\Delta Y \mid D, U, Z) \, d\Delta Y
} \\[1ex]
=& \frac{\Pr(R=0 \mid \Delta Y, D,U)p(\Delta Y \mid D, U, Z)}{\Pr(R=0 \mid D, U, Z)} \\[1ex]
=& p(\Delta Y \mid D, R = 0, U, Z)
\end{align*}

 Proof of Equation~(\ref{expectation_identification}).
\begin{align*}
\mathbb{E}[\Delta Y \mid D, R = 0, U, Z]
&= \int \Delta Y  p(\Delta Y \mid D, R = 0, U, Z) \, d\Delta Y \\[1ex]
&= \int \Delta Y  \frac{
  \OR(\Delta Y,D,U)  p(\Delta Y \mid D, R = 1, U, Z)
}{
  \mathbb{E}\left[ \OR(\Delta Y,D,U) \mid D, R = 1, U, Z \right]
} \, d\Delta Y \\[1ex]
&= \frac{
  \mathbb{E}\left[ \OR(\Delta Y,D,U)  \Delta Y \mid D, R = 1, U, Z \right]
}{
  \mathbb{E}\left[ \OR(\Delta Y,D,U) \mid D, R = 1, U, Z \right]
}
\end{align*}

\subsection{Identifiability of the Odds Ratio Model} \label{Appendix_OR_Iden}
Let $\widetilde{\OR}(\Delta Y, D, U) = \OR(\Delta Y, D, U)/\mathbb{E}\!\left[ \OR(\Delta Y, D, U) \mid D, R=1, U \right]$. Suppose there exists another square-integrable function $\widetilde{\OR}^*(\Delta Y, D, U)$ that also satisfies Equation~(\ref{OR_identifiability}); then
\[
\mathbb{E}\!\left[\widetilde{\OR}(\Delta Y, D, U)-\widetilde{\OR}^*(\Delta Y, D, U) \mid D, R=1, U, Z \right] = 0.
\]
If the completeness condition (Assumption~\ref{iden_assumption} (ii)) holds, it follows that $\widetilde{\OR}(\Delta Y, D, U)=\widetilde{\OR}^*(\Delta Y, D, U)$ almost surely, which guarantees uniqueness of the solution to Equation~(\ref{OR_identifiability}). Therefore, because $\OR(0,D,U)=1$, the odds ratio function is identified via $\OR(\Delta Y, D, U)= \widetilde{\OR}(\Delta Y, D, U) / \widetilde{\OR}(\Delta Y=0, D, U)$. This follows because
\begin{align*}
   \widetilde{\OR}(\Delta Y, D, U)= \frac{\Pr(R=0 \mid \Delta Y, D, U)/ \Pr(R=1 \mid \Delta Y, D, U)}{\Pr(R=0 \mid D, U)/ \Pr(R=1 \mid D, U)},
\end{align*}
and thus
\begin{align*}
   \frac{\widetilde{\OR}(\Delta Y, D, U)}{\widetilde{\OR}(\Delta Y=0, D, U)}
   = \frac{\Pr(R=0 \mid \Delta Y, D, U) \cdot \Pr(R=1 \mid \Delta Y=0, D, U)}{\Pr(R=1 \mid \Delta Y, D, U) \cdot \Pr(R=0 \mid \Delta Y=0, D, U)}= \OR(\Delta Y, D, U).
\end{align*}
The last equality follows from Equation~(\ref{OR_representation}) in Proposition~\ref{proposition_1}.

\subsection{Proof of Theorem~\ref{theorem_1}} 
\label{proof_theorem1}
Let $\delta=(\tau,\alpha,\beta,\gamma)$ and  $O=(R\Delta Y, D,R,U,Z)$. Define  $M(O; \delta)$ as the stacked estimating functions for $(\tau,\alpha,\beta,\gamma)$ described in Steps 3 and 4 of the Estimation subsection. Note that $\hat{\tau}$ solves the equation 
\begin{equation*}
  n^{-1}\sum_{i=1}^{n} \frac{R_i(D_i-\pi(U_i,Z_i;\hat{\beta}))}{\bar{D}q(\Delta Y_i,D_i,U_i;\hat{\alpha},\hat{\gamma})(1-\pi(U_i,Z_i;\hat{\beta}))}\Delta Y_{i}- \tau=0
\end{equation*}
 The specific form of $M(O; \delta)$ depends on the choice of $H(D,Z,U)$ and $G(U,Z)$, as well as on the specification of the working models. Suppose the assumptions in Theorem~\ref{theorem_1} hold. A Taylor expansion gives
\begin{equation*}
  0= n^{-1}\sum_{i=1}^{n} M(O_i;\delta^*) + \frac{\partial}{\partial \delta^\top} \mathbb{E}\left[M(O;\delta^*)\right](\hat{\delta}-\delta^*) + o_p(n^{-1/2})
\end{equation*}
Then, the theory of M-estimation ensures 
\begin{equation*}
   \sqrt{n}(\hat{\delta}-\delta^*) = -\left\{\frac{\partial}{\partial \delta^\top} \mathbb{E}\left[M(O;\delta^*)\right]\right\}^{-1} \frac{1}{\sqrt{n}}\sum_{i=1}^{n} M(O_i;\delta^*) + o_p(1) \xrightarrow{d} \mathcal{N}(0,V^*)
\end{equation*}
 where, letting $A=\mathbb{E}\left[\partial M(O;\delta^*)/\partial \delta^\top\right]$ and $B=\mathbb{E}\left[M(O;\delta^*)M(O;\delta^*)^\top\right]$, the sandwich variance is $V^*=A^{-1}B(A^{-1})^\top$.
Finally, the asymptotic variance of $\hat{\tau}$, $V_{\tau}$ corresponds to the relevant element of $V^*$. This completes the proof.

\end{document}